\documentstyle[emulateapj5,epsf]{aastex}

\def \farcs{\hbox{$.\!\!^{\prime\prime}$}}

\def \moverl{\hbox{${\rm M}_\odot/{\rm L}_{{\rm B}\odot}$}}

\lefthead{Hoekstra et al.}
\righthead{Weak lensing study of low mass groups}
\begin{document}

\slugcomment{Accepted for ApJ Letters}
\title{Weak lensing study of low mass galaxy groups: implications for 
$\Omega_m$ $^1$}

\author{H.~Hoekstra$^2$, M.~Franx$^3$, K.~Kuijken$^2$, R.G.~Carlberg$^{4,11}$,
	H.K.C~Yee$^{4,11}$, H.~Lin$^{4,5,6,11}$, S.L.~Morris$^{4,7,11}$, 
	P.B.~Hall$^{4,11}$, D.R.~Patton$^{4,8,11}$, M.~Sawicki$^{4,9,11}$, 
	and G.D.~Wirth$^{4,8,10,11}$}

\begin{abstract}

We report on the first measurement of the average mass and
mass-to-light ratio of galaxy groups by analysing the weak lensing
signal induced by these systems. The groups, which have velocity
dispersions of 50-400 km/s, have been selected from the Canadian
Network for Observational Cosmology Field Galaxy Redshift Survey
(CNOC2). This survey allows the identification of a large number of
groups with redshifts ranging from $z=0.12-0.55$, ideal for a weak
lensing analysis of their mass distribution. For our analysis we use a
sample of 50 groups which are selected on the basis of a careful
dynamical analysis of group candidates. We detect a signal at the 99\%
confidence limit. The best fit singular isothermal sphere model yields
an Einstein radius $r_E=0\farcs 72\pm 0\farcs29$. This corresponds to
a velocity dispersion of
$\langle\sigma^2\rangle^{1/2}=274^{+48}_{-59}$ km/s (using photometric
redshift distributions for the source galaxies), which is in good
agreement with the dynamical estimate. Under the assumption that the
light traces the mass, we find an average mass-to-light ratio of
$191\pm83~h\moverl$ in the restframe $B$ band. Unlike dynamical
estimates, this result is insensitive to problems associated with
determining group membership. After correction of the observed
mass-to-light ratio for luminosity evolution to $z=0$, we find
$254\pm110~h\moverl$, lower than what is found for rich clusters.  We
use the observed mass-to-light ratio to estimate the matter density of
the universe, for which we find $\Omega_m=0.19\pm0.10$
$(\Omega_\Lambda=0)$, in good agreement with other recent estimates.
For a closed universe $(\Omega_m+\Omega_\Lambda=1)$, we obtain
$\Omega_m=0.13\pm0.07$.

\end{abstract}


\section{Introduction}

Galaxy groups, like the Local Group, are common structures in the
universe.  Although being numerous, groups are difficult to identify
because the contrast with the smooth background of galaxies is quite
low, and their galaxy properties are similar to that of the field. To
date most systems have been studied using the results of large
redshift surveys (e.g., Turner \& Gott 1976; Ramella, Geller, \&
Huchra 1989; Huchra, Geller, \& Corwin 1995) or X-ray observations
(Mulchaey et al. 1996). Unfortunately, X-ray observations are only available
for very few groups.

\vbox{
\vspace{0.3cm}
\footnotesize
\noindent $^1$~Based on observations made with the William Herschel
Telescope operated on the island of La Palma by the Isaac Newton Group
in the Spanish Observatorio del Roque de los Muchachos of the
Instituto de Astrofisica de Canarias.\\
$^2$~Kapteyn Astronomical Institute, University of
Groningen, P.O.~Box 800, 9700 AV Groningen, The Netherlands;
hoekstra,kuijken@astro.rug.nl\\
$^3$~Leiden Observatory, P.O.~Box 9513, 2300 RA Leiden,
The Netherlands; franx@strw.strw.leidenuniv.nl\\
$^4$~Department of Astronomy, University of Toronto,
Toronto, Ontario M5S 3H8, Canada; hyee, carlberg, hall,
patton@astro.utoronto.ca\\
$^5$~Present address: Steward Observatory, University of
Arizona, Tucson, AZ 85721; hlin@as.arizona.edu\\
$^6$~Hubble Fellow\\
$^7$~Dominion Astrophysical Observatory, Herzberg
Institute of Astrophysics, National Research Council, 5071 W. Sannich
Rd., Victoria, BC V8X 4M6, Canada; Simon.Morris@hia.nrc.ca\\
$^8$~Department of Physics and Astronomy, University of
Victoria, Victoria, BC, V8W 3P6, Canada\\
$^9$~Present address: CalTech, Mail Code 320-47, Pasadena,
CA 91125; sawicki@pirx.caltech.edu\\
$^{10}$~Present address: Keck Observatory, Waimea, HI 96743;
wirth@keck.hawaii.edu\\
$^{11}$~Visiting Astronomer, Canada-France-Hawaii Telescope, which
is operated by the National Research Council of Canada, Le Centre 
National de Recherche Scientifique, and the University of Hawaii
}

Measuring the amount of matter locked up in these typical systems is
interesting, but a measurement of the average mass-to-light ratio of
galaxy groups may be even more important as it provides a good measure
of the M/L ratio of the field, i.e., the universe as a whole
(e.g., Gott \& Turner 1977).  Subsequently, this result can be used to
obtain an estimate for the matter density $\Omega_m$ (Oort 1958; Gott
\& Turner 1977), similar to what has been done for rich clusters of
galaxies (e.g., Carlberg et al. 1997; Carlberg et al. 1999).

However, measuring the mass or M/L ratio of groups selected
from redshift surveys is difficult. Nolthenius \& White (1987) showed
that the dynamical masses inferred from such surveys depend on the
survey parameters, the group selection procedure, and the way galaxies
cluster. Consequently, an independent measure of the group mass is
needed. In this letter we study galaxy groups by their weak lensing
effect on the shapes of the images of the faint background sources.

Weak lensing enables us to measure the projected mass surface density,
without any assumption about the geometry or dynamical state of the
system under investigation. The technique has been applied successfully
to rich clusters of galaxies (for a review see Mellier 1999).

The amplitude of the weak lensing signal is proportional to the 
mass of the lens, and as a result the expected signal from an
individual galaxy group is very low. To circumvent this problem we
study the properties of the ensemble averaged group by stacking
the signals of many groups at intermediate redshifts, where the lensing
signal is maximal.

Galaxy groups identified in the Canadian Network for Observational
Cosmology Field Galaxy Redshift Survey (CNOC2) (Lin et al.  1999; Yee
et al. 2000; Carlberg et al. 2000) are ideal for a weak lensing study
of their mass distribution: the survey provides a large sample of
groups at intermediate redshifts, which can be imaged efficiently by
wide field imaging.

The observations, data reduction and weak lensing analysis are
outlined in \S~2, and the group selection is discussed in \S~3. The
results of the weak lensing analysis are given in \S~4. In \S~5 we
present our estimate of $\Omega_m$.

\section{Observations and analysis}

The CNOC2 survey targeted four widely separated patches on the sky to
study the dynamics of galaxy clustering at intermediate redshifts.
Redshifts of approximately 5000 galaxies down to $R_C=21.5$ have
been measured, resulting in a large sample of galaxies at intermediate
redshifts $(z=0.12-0.55)$. A detailed description of the
survey and the catalogues is given in Yee et al. (2000).

We have observed the central parts of the two patches 1447+09 and 2148-05
using the 4.2m William Herschel Telescope at La Palma. The data were
taken using the prime focus camera, equipped with a thinned 2k$\times$4k
pixel EEV10 chip, and a pixel scale of $0\farcs 237$ pixel$^{-1}$.
To cover the central regions of the patches, a mosaic of 6 pointings
was observed, resulting in a field of view of 31 by 23 arcminutes. 

The typical total integration time per pointing is 1 hour in $R$. The
seeing ranged from $0\farcs6$ to $1\farcs0$, with a median seeing of
$0\farcs7$ for the 1447 field and $0\farcs85$ for the 2148 field.  The
images were debiased and flatfielded, and photometric calibration was
performed using standard stars from Landolt (1992). 

Our object analysis is based on the procedure developed by Kaiser,
Squires, \& Broadhurst (1995) and Luppino \& Kaiser (1997), with a
number of modifications which are described in Hoekstra et al. (1998)
and Hoekstra, Franx, \& Kuijken (2000). The objects detected in the
images are analysed: sizes, apparent magnitudes, and shape parameters
are determined. We correct the measurements for various observational
distortions. We also tested how uncertainties in these corrections
would affect our results, and found that the results are very stable.

In the weak lensing analysis we use galaxies with $22<R<26$ as
background sources. These catalogs contain some faint group members,
which can lower the lensing signal at the group centres. We examined
the average number density of sources around the groups, and found
that the contamination is negligible.

\section{Galaxy Groups}

Finding galaxy groups in a redshift survey such as CNOC2 is a
difficult problem. The crucial step is to determine the group
membership, which makes a reliable dynamical mass estimate
difficult. The weak lensing mass estimate is more robust against
contamination by interlopers, as it relies only on the position of the
overdensity.

Lensing in itself does not provide information as to whether the studied
structures are gravitationally bound galaxy groups. The question
whether the selected structures are genuine groups is less important
when determining the average M/L ratio of these systems.  As
we will demonstrate in \S~4.3 the M/L ratio from weak
lensing is particularly stable against uncertainties in the
determination of group membership. For the estimate of $\Omega_m$
presented in \S~5 it is sufficient to identify high density regions.

For our analysis we use the groups presented by Carlberg et
al. (2000). The groups are found using an iterative method,
which is a variant of the often used friends-of-friends
algorithm. A detailed discussion of the algorithm is
presented in Carlberg et al. (2000).

The resulting sample consists of 50 groups which are within the observed
fields. The average redshift of the groups is $z=0.33$, and
they have velocity dispersions ranging from $50-400$ km/s.
The redshift information is crucial because the contrast of the groups
with the field is low: the average group corresponds to a 1.2$\sigma$
overdensity in number counts. 

To estimate the light contents of the groups, we determine $B_z$, the
restframe $B$-band magnitude. To this end, we use template spectra
for a range in spectral types and compute the corresponding passband
corrections as a function of redshift and galaxy colour. The redshifts
and colours of the galaxies are taken from the CNOC2 catalogues (e.g., Yee
et al. 2000). 

To account for the incompleteness of the redshift survey, each galaxy
with a measured redshift is assigned a proper weight (Yee et al. 2000;
Lin et al. 1999). These weights are used to correct the galaxy
luminosities, and the group luminosity profiles (the average
correction factor is found to be $1.53\pm0.02$). We used the
luminosity function derived by Lin et al. (1999) to estimate the
correction factor for the missing faint galaxies, for which we find a
value of $1.20\pm0.07$.  We find that the average luminosity of the
groups considered here is $L_B=(6.3\pm0.6)\times 10^{10}h^{-2}{\rm
L}_{B\odot}$, which corresponds to $\sim 5 L_{B\star}$.  More details
on the light distribution of the groups can be found in Carlberg et
al.~(2000).

\section{Mass and mass-to-light of galaxy groups}

\subsection{Weak lensing signal}

We stack the average distortion as a function of radius around the
groups, because the number density of sources ($\sim 35$
arcmin$^{-2}$) is too low to detect a significant signal for
individual groups. The amplitude of the lensing signal depends on the
redshift of the group, and we scale the contribution of each group to
the average signal such that it corresponds to that of the `average'
group at a redshift $z=0.33$ (see \S 4.2).

We quantify the lensing signal by means of the tangential distortion,
which is defined as $g_T=-(g_1 \cos 2\phi + g_2 \sin 2\phi)$, where
$\phi$ is the azimuthal angle with respect to the assumed group
centre, and $g_i$ are the components of the distortion. The number of
confirmed group members per group is rather low ($3-7$ members), and
therefore the positions of the group centres are somewhat uncertain.
Here we use the positions which are found from a virial analysis
described in Carlberg et al. (2000). Simulations using SIS models
indicate that the uncertainty in the group centres could result in an
overestimate of the average lensing signal by at most 4 to 8\%. Also,
the groups are embedded in the large scale structure which tends to
increase the weak lensing mass estimate. The numerical results from
Cen (1997) match well our observations, and indicate that the bias is
still small, on the order of 10\%. We note, however, that our estimate
for the average M/L ratio presented in \S 4.3 does not
suffer from either of the biases mentioned above.

The azimuthally averaged tangential distortion around the 50 groups is
presented in Figure~\ref{gtprof}a. We detect a positive average
tangential distortion of $0.00254\pm 0.0011$, which is significant at
the 99\% confidence level. We tested the robustness of the signal in
various ways.  Figure~\ref{gtprof}b shows the result when the phase of
the distortion is increased by $\pi/2$.  Also other tests, such as
randomizing the group positions or the ellipticities of the sources
yielded no significant signal. We therefore conclude that the observed
signal is due to weak lensing by galaxy groups.

The best fit singular isothermal sphere model $(\kappa(r)=r_E/2r)$ to
the ensemble averaged distortion from the sample of 50 galaxy groups from 
Carlberg et al. (2000) yields an Einstein radius of $r_E=0\farcs72\pm
0\farcs29$. 

\subsection{Estimate of the velocity dispersion}

The next step is to relate the Einstein radius to an estimate of the
velocity dispersion. To do so we use photometric redshift
distributions inferred from the Hubble Deep Fields
(Fern{\'a}ndez-Soto, Lanzetta, \& Yahil 1999; Chen et al. 1998), which
generally work well (Hoekstra et al. 2000).  We determine the $R$ band
magnitudes of the galaxies in the HDFs, and use these results to
derive the average group velocity dispersion.
\vbox{
\begin{center}
\leavevmode 
\hbox{%
\epsfxsize=0.97\hsize 
\epsffile[20 175 570 690]{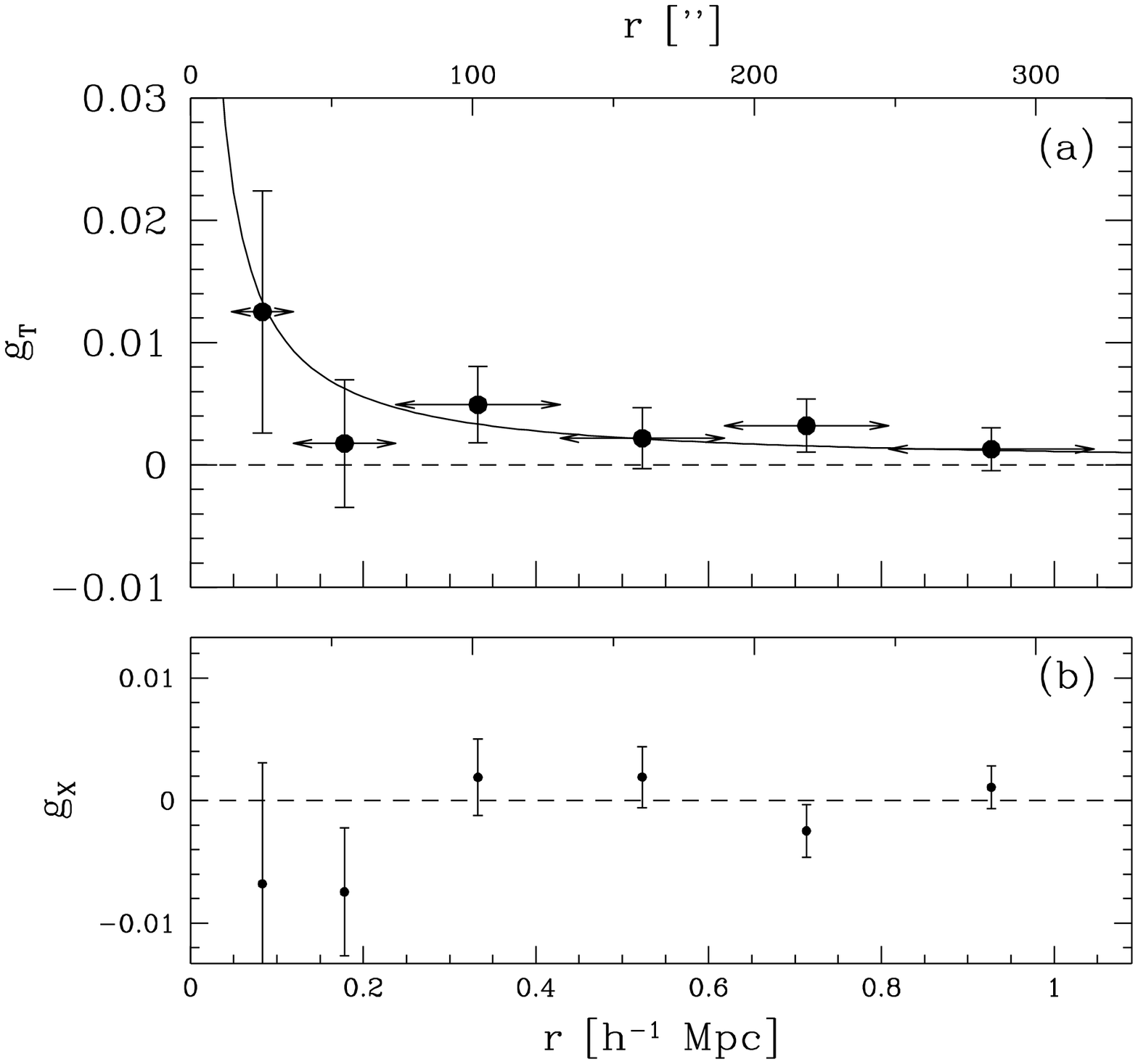}} 
\figcaption{\footnotesize (a) The ensemble averaged tangential distortion
as a function of radius around the 50 galaxy groups from Carlberg et
al. (2000).  The amplitude of the signal corresponds to that of the
`average' group at the median group redshift $z=0.33$. The errorbars
are determined as described in Hoekstra et al. (2000) and indicate the
uncertainty in the measurements of the shapes of the sources. The
profile of the best fit singular isothermal sphere model (to the solid
points), which has a velocity dispersion of $274^{+48}_{-59}$ km/s, is
indicated by the solid line. (b) The signal when the phase of the
distortion is increased by $\pi/2$: no signal should be present if the
signal in (a) is due to lensing. The vertical errorbars indicate the
$1\sigma$ errors\label{gtprof}}
\end{center}}

\vbox{
\begin{center}
\leavevmode 
\hbox{%
\epsfxsize=0.97\hsize
\epsffile[70 155 585 690]{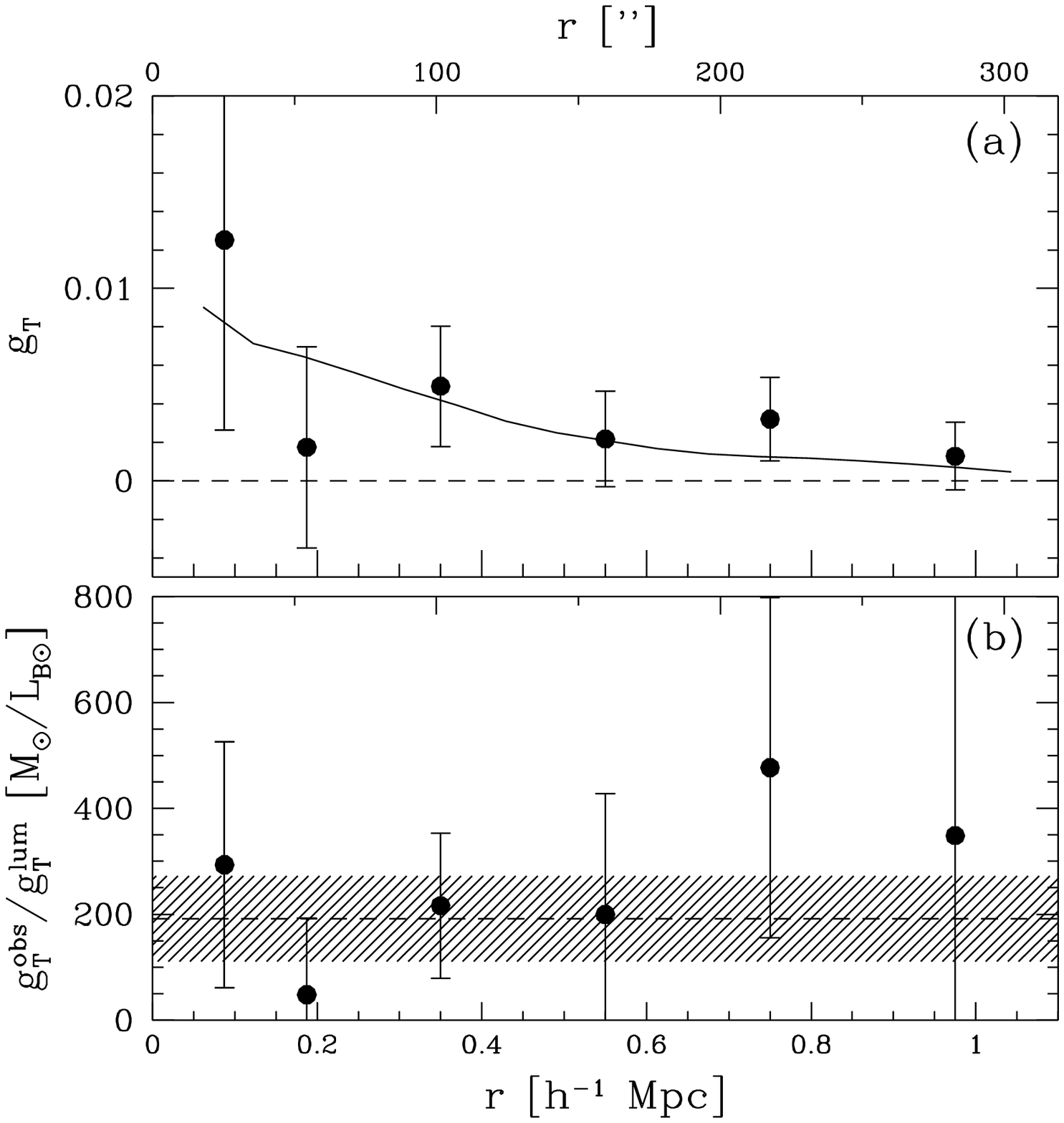}}
\figcaption{\footnotesize (a) Plot of the average tangential distortion
as a function of radius from the ensemble of 50 galaxy groups from the
CNOC2 survey.  The solid line is the expected tangential distortion
(scaled by the M/L ratio to fit the observations) derived
from the average radial light profile, under the assumption that the
M/L ratio is constant with radius. (b) The ratio of the
observed distortion and the derived distortion from the light (taking
$M/L_B=1$ in solar units). The shaded region indicates the one
$\sigma$ region around the weighted average of the points. The
observations are consistent with a constant M/L ratio of
$191\pm83~h\moverl$.
\label{moverl}}
\end{center}}

The strength of the lensing signal is characterized by $\beta$, which
is defined as $\beta=\max[0,D_{ls} /D_s]$, where $D_{ls}$ and $D_s$
are the angular diameter distances between the lens and the source,
and the observer and the source. For each group-galaxy pair we compute
the corresponding value of $\beta$ based on the $R$ band magnitude of
the source and the redshift of the group.  We find
$\langle\beta\rangle=0.393\pm0.006$, which results in an ensemble
averaged group velocity dispersion of
$\langle\sigma^2\rangle^{1/2}=274^{+48}_{-59}$ km/s (68\% confidence;
$\Omega_m=0.2$, and $\Omega_\Lambda=0$) for the sample of groups from
Carlberg et al. (2000).  For $\Omega_m=0.2$ and $\Omega_\Lambda=0.8$
it changes to $\langle\sigma^2\rangle^{1/2}=258^{+45}_{-56}$ km/s
(68\% confidence). These results are in good agreement with the
average velocity dispersion of 230 km/s from the velocities of the
group members.

\subsection{Mass-to-light ratio}

Under the assumption that the light traces the mass, we derive the
expected tangential distortion as a function of radius.  We use the
ensemble averaged group luminosity profile to calculate the expected
signal for each group for a mass-to-light ratio of unity (taking into
account the redshift), and average the result for all groups. To
measure the M/L ratio, we scale the resulting tangential
distortion to match the observed signal. 

In Figure~\ref{moverl}a the resulting profile (solid line) is
shown. The ratio of the computed and observed signal is presented in
Figure~\ref{moverl}b, and is consistent with a constant M/L
ratio with radius for which we find a value of $191\pm83~h\moverl$ in
the restframe $B$ band.

This measurement of the M/L ratio has not been corrected for
luminosity evolution. If the luminosity evolution scales with redshift
as $L_B\propto(1+z)$ (e.g., Lin et al. 1999), we obtain a value of
$254\pm110~h\moverl$, corrected to $z=0$. Carlberg et al. (1997)
measured the M/L ratio of a sample of 15 rich clusters, for
which they found an average value of $M/L_r=237 \pm 41 {\rm
M}_\odot/{\rm L}_{r\odot}$.  To convert their result to a
M/L ratio in the $B$ band, we assume an average colour of
the cluster of $B-r=1.07$, which corresponds to the typical colour of
S0 galaxies (J{\o}rgensen et al. 1995).  Thus we find that the average
cluster M/L ratio is $438\pm76~h\moverl$ (where we also
corrected for luminosity evolution to $z=0$). Thus the average group
M/L ratio in the $B$ band is lower than the value typically
found for rich clusters.

We derived the expected lensing signal using only galaxies identified
as group members. However, lensing is sensitive to the contribution of
all matter along the line of sight.  To examine the contribution of
the remaining galaxies we recalculated the group light profile, using
all galaxies with redshifts.  Fitting the predicted distortion to the
observations, we find a M/L ratio of $183\pm80~h\moverl$, in excellent
agreement with our measurement from group members only. An important
consequence of this exercise is that the weak lensing estimate of the
M/L ratio is insensitive to the determination of group
membership, unlike the dynamical estimators.
  
\section{Estimate of $\Omega_m$}

A well known method to estimate the matter density of the universe was
proposed by Oort (1958): $\Omega_m$ is the product of the universe's
mass-to-light ratio and its luminosity density. Carlberg et al. (1997)
used the observed M/L ratios of a sample of rich clusters 
to estimate $\Omega_m$, for which they found a value of $\Omega_m=0.19\pm0.06$.

The galaxy properties of rich clusters are quite different from that
of the field, and a large correction is needed to relate the cluster
M/L ratio to the M/L ratio of the
universe. However, we found a small difference of
$\Delta(B-V)=0.035\pm0.013$ between the average restframe colours of
group galaxies and the field, which is caused by a small difference in
stellar populations. We use stellar evolution models\footnote{We used
the latest models (1999), which can be obtained from {\it ftp.iap.fr}
in {\it pub/from\_users/charlot}} (Bruzual \& Charlot 1993) to make a
small correction for this effect. Under the assumption that the fraction
of the mass in stars is the same for both groups and galaxies, we
find that the $B$ band M/L ratio of the field is lower by a 
factor 1.15, compared to the value found for the groups.

We combine our estimate of the M/L ratio with the
results from Lin et al. (1999), which are based on the same
data. Convolving the redshift distribution of the groups with their
redshift dependent luminosity density yields $j=(3.2\pm0.6)\times 10^8
h {\rm L}_{B\odot} {\rm Mpc}^{-3}$ (assuming $\Omega_m=0.2$ and
$\Omega_\Lambda=0$).

We obtain $\Omega_m=0.19\pm0.10$ for an $\Omega_\Lambda=0$ cosmology.
Our estimate for $\Omega_m$ decreases to a value of
$\Omega_m=0.13\pm0.07$ for $\Omega_\Lambda=0.87$. The value for
$\Omega_m$ is derived in a self consistent way, and therefore is
independent of the cosmology assumed throughout the paper. The error
is dominated by the uncertainty in the weak lensing signal due to the
intrinsic ellipticities of the sources, but also incorporates the
uncertainties in the determination of the luminosity density and the
group luminosities. The systematic uncertainty $(\sim 15\%)$
introduced by the colour difference between the group members and the
field is not included in this error estimate.

Our results on $\Omega_m$ agree well with the result from Carlberg et
al. (1997), and combined constraints from high redshift supernovae
(e.g., Perlmutter et al. 1999) and CMB measurements (e.g., Efstathiou
et al. 1999, De Bernardis et al. 2000)

Some caveats should be noted as well. Our measurement of the
M/L ratio is stable against changes in group membership, but
is only correct if the light traces the mass. If the dark matter is
more extended than the light our estimate for $\Omega_m$ should be
interpreted as a lower limit. Also the correction for the colour
difference between group members and the field is somewhat uncertain.

\acknowledgments

We thank Jocelyn B{\'e}zecourt for his help to obtain part of the
imaging data. The WHT observations for this project have been
supported financially by the European Commission through the TMR
program `Access to large-scale facilities', awarded to the Instituto
de Astrofisica de Canarias. HH acknowledges support from the Leidsch
Kerkhoven-Bosscha Fonds and the University of Toronto.


\begin{thebibliography}{}
\bibitem{BC93}
	Bruzual, G., \& Charlot, S. 1993, ApJ, 405, 538
\bibitem{C97}  
        Carlberg, R.G., Yee, H.K.C., \& Ellingson, E. 1997, ApJ, 478, 462
\bibitem{C99}
	Carlberg, R.G., Yee, H.K.C., Morris, S.L., Lin, H., Ellingson, E.
	Patton, D., Sawicki, M., \& Shepherd, C.W. 1999, ApJ, 516, 552
\bibitem{C00}
	Carlberg, R.G., Yee, H.K.C., Morris, S.L., Lin, H., Hall, P.B.,
	Patton, D.R., Sawicki, M., \& Shepherd, C.W. 2000, ApJ submitted,
	(astro-ph/0008201)
\bibitem{Cen97}
	Cen, R. 1997, ApJ, 485, 39
\bibitem{Cetal98}
        Chen,~H.-W., Fern{\'a}ndez-Soto,~A., Lanzetta, K.M., Pascarelle, S.M.,
        Puetter, R.C., Yahata, N., \& Yahil, A., preprint, astro-ph/9812339
\bibitem{DB00}
	De Bernardis, P., et al. 2000, Nature, 404, 955	
\bibitem{E99}
	Efstathiou, G., Bridle, S.L., Lasenby, A.N., Hobson, M.P., 
	\& Ellis, R.S. 1999, MNRAS, 303, 47
\bibitem{FLY99}
        Fern{\'a}ndez-Soto, A., Lanzetta, K.M., \& Yahil, A. 1999, ApJ, 513, 34
\bibitem{GT77}
	Gott, J.R. III, \& Turner, E.L. 1977, ApJ, 213, 309
\bibitem{HFKS98}
	Hoekstra, H., Franx, M., Kuijken, K., \& Squires, G. 1998, ApJ,
	504, 636
\bibitem{HFK00}
	Hoekstra, H., Franx, M., Kuijken, K. 2000, ApJ, 532, 88
\bibitem{HGC95}
	Huchra, J.P., Geller, M.J., \& Corwin, H.G. 1995, ApJS, 99, 391
\bibitem{J95}
	J{\o}rgensen, I., Franx, M., \& Kj{\ae}rgaard, P. 1995, 
	MNRAS, 273, 1097
\bibitem{ksb95} 
	Kaiser, N., Squires, G., \& Broadhurst, T. 1995, ApJ, 449, 460
\bibitem{L92}
	Landolt, A.U. 1992, AJ, 104, 340
\bibitem{L99}
	Lin, H., Yee, H.K.C., Carlberg, R.G., Morris, S.L., Sawicky, M.,
	Patton, D.R., Wirth, G., \& Shepherd, C.W. 1999, ApJ, 518, 533 
\bibitem{LK97}
        Luppino, G.A., Kaiser, N. 1997, ApJ, 475, 20
\bibitem{M99}
        Mellier, Y. 1999, ARA\&A, 37, 127
\bibitem{M96}
	Mulchaey, J.S., Davis, D.S., Mushotzky, R.F., \& Burstein, D. 
	1996, 456, 80
\bibitem{NW87}
	Nolthenius, R., \& White, S.D.M. 1987, MNRAS, 235, 505
\bibitem{O58}
	Oort, J.H. 1958, in La Structure et L'{\'e}volution de L'Universe,
	Onzi{\`e}me Conseil de Physique, ed. R. Stoops (Solvay: Bruxelles),
	163
\bibitem{P99}
	Perlmutter, S., et al. 1999, ApJ, 517, 565
\bibitem{R89}
	Ramella, M., Geller, M.J., \& Huchra, J. 1989, ApJ, 344, 57
\bibitem{TG76}
	Turner, E.L., \& Gott, J.R. III 1976, ApJS, 32, 409
\bibitem{Y96}
	Yee, H.K.C., Ellingson, E., \& Carlberg, R.G. 1996, ApJ, 102, 269
\bibitem{Y00}
	Yee, H.K.C. et al. 2000, ApJS, in press, astro-ph/0004026

\end{thebibliography}
\end{document}